\documentclass[11pt]{article}
\usepackage{latexsym}  
\usepackage{amssymb}
\usepackage{graphicx}
\usepackage{amsmath}

\begin{document}

\begin{center}
{\Large\bf Charged Lepton Mass Relations } \\[.1in]
{\Large\bf in a Supersymmetric Yukawaon Model}

\vspace{5mm}
{\bf Yoshio Koide}

{\it IHERP, Osaka University, 1-16 Machikaneyama, 
Toyonaka, Osaka 560-0043, Japan} \\
{\it E-mail address: koide@het.phys.sci.osaka-u.ac.jp}

\date{\today}
\end{center}

\vspace{3mm}
\begin{abstract}
According to an idea that effective Yukawa coupling
constants $Y_f^{eff}$ are given vacuum expectation
values $\langle Y_f\rangle$ of fields (``Yukawaons") 
$Y_f$  as $Y_f^{eff}=y_f \langle Y_f\rangle/\Lambda$,
a possible superpotential form in the charged lepton
sector under a U(3) [or O(3)] flavor symmetry 
is investigated.
It is found that a specific form of the superpotential 
can lead to an empirical charged lepton mass 
relation without any adjustable parameters.
\end{abstract}

\vspace{3mm}

\noindent{\large\bf 1 \ Introduction}

The so-called ``Yukawaon model" [for example, see Ref.\cite{Yukawaon}]  
claims that, in effective Yukawa interactions 
of quarks and leptons 
$$
H_Y = \sum_{i,j} \ell_i (Y_e^{eff})_{ij} e_j^c H_d +\cdots ,
\eqno(1.1)
$$
the effective Yukawa coupling constants $Y_f^{eff}$ 
($f=e,\nu,u,d$) are
given by the vacuum expectation values (VEVs) 
$\langle Y_f \rangle$ of a scalar field $Y_f$ as
$$
Y_f^{eff} = \frac{y_f}{\Lambda}  \langle Y_f \rangle .
\eqno(1.2)$$
Here, for simplicity, we have explicitly denoted only the
charged lepton sector.
In Eq.(1.1), $\ell$ and $e^c$ are SU(2)$_L$ doublet and singlet
fields, respectively, and $\Lambda$ is an energy scale of 
the effective theory. 
(We have considered a supersymmetric (SUSY) scenario.) 
Hereafter, we refer the fields $Y_f$ as ``Yukawaons"
\cite{Yukawaon}, which are gauge singlets.
In addition to the Yukawaon $Y_e$, we consider
a field $\Phi_e$ which is related to $Y_e$ as
$$
\langle Y_e \rangle = k \langle \Phi_e \rangle
 \langle \Phi_e \rangle .
 \eqno(1.3)
$$
We also refer $\Phi_e$ as  a ``ur-Yukawaon",
which has been introduced in order to fix the VEVs of 
the Yukawaon $Y_e$.
(For the moment, we consider the ur-Yukawaon only in the
charged lepton sector.)
Then, an empirical charged lepton mass formula \cite{Koidemass}
$$
R_e \equiv  \frac{m_e +m_\mu + m_\tau}{
(\sqrt{m_e} +\sqrt{m_\mu} + \sqrt{m_\tau})^2 } 
=\frac{2}{3} ,
\eqno(1.4)
$$
is rewritten as
$$
R_e \equiv  \frac{v_1^2 +v_2^2 +v_3^2}{
(v_1+v_2+v_3)^2 } 
=\frac{2}{3} ,
\eqno(1.5)
$$
where $v_i = \langle (\Phi_e)_{ii} \rangle$.

Previously, the author \cite{K-mass90} has derived
the relation (1.5) by assuming the following 
U(3)-flavor-invariant scalar potential
$$
V= \mu^2 (\pi^2 + \eta^2 + \sigma^2) 
+\lambda (\pi^2 + \eta^2 + \sigma^2)^2
+ \lambda'  (\pi^2 + \eta^2) \sigma^2 ,
\eqno(1.6)
$$
where
$$
\pi=\frac{1}{\sqrt2}(\Phi_{11} -\Phi_{22}) ,
\ \ \eta =\frac{1}{\sqrt6}(\Phi_{11} +\Phi_{22}-2\Phi_{33}) ,
\ \ \sigma =\frac{1}{\sqrt3}(\Phi_{11} +\Phi_{22}+\Phi_{33}) ,
\eqno(1.7)
$$
and $\pi^2 + \eta^2 + \sigma^2$ and $\sigma^2$ correspond to
${\rm Tr}[\Phi\Phi]$ and $\frac{1}{3}{\rm Tr}^2[\Phi]$
in a diagonal basis of the VEV matrix $\langle \Phi \rangle$,
respectively.
Here, we have dropped the index ``$e$"  in $\Phi_e$
for convenience.  
Since, in the present paper, we often meet with traces of 
matrices $A$,  hereafter, 
we denote the traces  ${\rm Tr}[A]$ as $[A]$ concisely. 
The scalar potential (1.6) can be rewritten as
$$
V=\mu^2 [\Phi\Phi] +\lambda [\Phi\Phi]^2 +\frac{1}{3}\lambda'
[\Phi^{(8)} \Phi^{(8)}][\Phi]^2 ,
\eqno(1.8)
$$
where $\Phi^{(8)}$ is an octet part of the nonet field
$\Phi$, $\Phi^{(8)} =\Phi -\frac{1}{3} [\Phi]$.
The minimizing condition\footnote{
The stability condition 
$\partial^2 V/\partial \Phi \partial \Phi >0$
will put a further constraint for a value of $\lambda'/\lambda$.
}
 of $V$ demands
$$
\frac{\partial V}{\partial \Phi}  =
2\left(\mu^2 +2 \lambda [\Phi\Phi]
+\frac{1}{3}\lambda'  [\Phi]^2 \right) \Phi +
\frac{2}{3}\lambda' \left( [\Phi\Phi] -\frac{2}{3}[\Phi]^2
\right) [\Phi] = 0 ,
\eqno(1.9)
$$
so that we obtain the relation (1.5), i.e.
$$
R=\frac{[\Phi\Phi]}{[\Phi]^2} = \frac{2}{3} ,
\eqno(1.10)
$$
together with 
$\mu^2+2\lambda [\Phi\Phi] +\frac{1}{3}\lambda'
[\Phi]=0$.\footnote{
The condition (1.9) has a form $c_1 \langle\Phi\rangle
+c_0 {\bf 1} =0$, so that the eigenvalues $v_i$ of 
$\langle\Phi\rangle$ are given by $v_i=-c_0/c_1$.
Since we know that the observed charged lepton masses
$m_{ei}$ have nonzero and nondegenerate values, we
want a set of $(v_1,v_2,v_3)$ with nonzero and 
nondegenerate values.
Then, we must require $c_1=c_0=0$, so that we obtain
the result (1.10).
}
Of course, a statement that 
the relation (1.10) was derived by assuming U(3) symmetry
is not correct.
The accurate statement is that the relation (1.10) was
derived from a scalar potential (1.6) [(1.8)] which is 
invariant under U(3) symmetry, but which is not a general 
form of the U(3) invariant scalar potential.

A straightforward SUSY version of the scalar potential 
(1.8) is as follows:
the superpotential $W$ is given by
$$
W= \mu [\Phi A] +\lambda' [\Phi] [\Phi^{(8)} B],
\eqno(1.11)
$$
where $A$ and $B$ are additional nonet fields.
Then, the superpotential (1.11) leads to a scalar potential
$$
V= |\mu|^2 [\Phi\Phi^\dagger]
+|\lambda'|^2 [\Phi] [\Phi]^\dagger
[\Phi^{(8)} \Phi^{(8)\dagger}] + \cdots.
\eqno(1.12)
$$
However, although the minimizing condition of the scalar potential 
(1.12)
can lead to the relation (1.10), the vacuum is not stable,
because there is another lower vacuum $V=0$ (a SUSY vacuum) at 
$\langle\Phi\rangle =0$.

A supersymmetric approach with SUSY vacuum conditions
to the mass relation (1.4)
has first been done by Ma \cite{K-mass-Ma}. 
His model with a flavor symmetry $\Sigma(81)$ is impeccable, 
but somewhat intricate.
Stimulated by his work, the author \cite{Koide-JHEP07} has also 
proposed a superpotential with a simple form
$$
W= \mu [\Phi \Phi] + \lambda  [\Phi \Phi \Phi],
\eqno(1.13)
$$
by assuming a Z$_2$ symmetry in addition to the U(3) flavor symmetry.
Here, it has been assumed that the octet
$\Phi^{(8)}=\Phi -\frac{1}{3}[\Phi]$ and singlet 
 $\Phi^{(1)}=\frac{1}{3}[\Phi]$
have Z$_2$ parities $-1$ and $+1$, respectively.
Then, in the cubic term
$$
[\Phi \Phi \Phi]=[\Phi^{(8)} \Phi^{(8)} \Phi^{(8)}]
+[\Phi^{(8)} \Phi^{(8)} ] [\Phi]
+\frac{1}{3} [\Phi^{(8)}] [\Phi]^2
+\frac{1}{9} [\Phi]^3 ,
\eqno(1.14)
$$
the Z$_2$ parities of the terms 
$[\Phi^{(8)} \Phi^{(8)} \Phi^{(8)}]$ and 
$[\Phi^{(8)}][\Phi]^2$ are $-1$,
so that those terms are dropped under the Z$_2$
symmetry:
$$
[\Phi \Phi \Phi]_{Z_2=+1} =
[\Phi^{(8)} \Phi^{(8)} ] [\Phi]+\frac{1}{9} [\Phi]^3
=[\Phi][\Phi \Phi] -\frac{2}{9} [\Phi]^3.
\eqno(1.15)
$$
Then, by requiring a SUSY vacuum condition 
$$
\frac{\partial W}{\partial \Phi} = 
2 (\mu+\lambda [\Phi] )\Phi +
\lambda \left( [\Phi\Phi]-\frac{2}{3} [\Phi]^2\right)
[\Phi] =0 ,
\eqno(1.16)
$$
where we have used 
$[\Phi^{(8)} \Phi^{(8)} ]=[\Phi\Phi] -\frac{1}{3} [\Phi]^2$, 
we can obtain the relation (1.10).
However, such the Z$_2$ charge assignment requires a somewhat
intricate scenario \cite{Koide-JHEP07} when $\Phi$ is related to
$Y$, because we need not only 
$\Phi^{(8)} \Phi^{(8)}+\Phi^{(1)} \Phi^{(1)}$ with $Z_2=+1$,
but also $\Phi^{(8)} \Phi^{(1)}+\Phi^{(1)} \Phi^{(8)}$
with $Z_2=-1$ in $Y=k \Phi \Phi$.

If we accept a higher dimensional term in the superpotential,
by assuming a simple form without such Z$_2$ symmetry 
$$
W= \mu [\Phi \Phi] + \frac{1}{\Lambda} [\Phi]^2
[\Phi^{(8)} \Phi^{(8)}],
\eqno(1.17)
$$
we can also obtain the relation (1.10):
$$
\frac{\partial W}{\partial \Phi} =2 \left( \mu +
\frac{1}{\Lambda}[\Phi]^2 \right) \Phi +
\frac{2}{\Lambda}
\left( [\Phi \Phi] -\frac{2}{3} [\Phi]^2 {\bf 1}
\right) [\Phi] =0 .
\eqno(1.18)
$$
However, we must recall that each Yukawaon $Y_f$ has a different
U(1)$_X$ charge $Q_X=x_f$ in order to distinguish each fermion
partner \cite{Koide-O3-PLB08}. 
Since the ur-Yukawaon $\Phi_e$ also has a U(1)$_X$ charge
$Q_X=\frac{1}{2}x_e$,
we cannot write the superpotential (1.17) [also Eq.(1.13)] 
without violating the U(1)$_X$ symmetry.

We would like to search for a superpotential form whose vacuum 
conditions lead to the relation (1.10) under the conditions
that (i) the superpotential $W$ does not include a higher
dimensional term, and  (ii) $W$ is invariant under the U(3)
[or O(3)] and U(1)$_X$ symmetries. 
Note that, in the original idea (1.6), the result (1.10) is 
obtained independently of the explicit parameter values $\mu$,
$\lambda$ and $\lambda'$.
We consider that such a motive should 
be inherited in a SUSY version of the scenario, too.
The result (1.10) should be obtained without adjusting parameters
in the model.
We will search for a superpotential form by considering that 
the form may include an ad hoc term for the time being, 
but the form should be simple.

\vspace{3mm}

\noindent{\large\bf 2 \ Ansatz and VEV relations}

In the present paper, we assume the Yukawaons $Y_f$ are
nonets of a U(3)-flavor symmetry [or ${\bf 5}+{\bf 1}$ of
O(3)], and those do not solely appear 
as octets of U(3) [or ${\bf 5}$-plets of O(3)] in the superpotential.
On the other hand,  as suggested 
by the forms (1.11) and (1.17), the traceless part of $\Phi_e$, 
$\hat{\Phi}_e \equiv \Phi_e -\frac{1}{3} [\Phi_e]$, seems to
play an crucial role in obtaining the relation (1.10). 
Therefore, for the ur-Yukawaon $\Phi_e$, we consider that
the traceless part $\hat{\Phi}_e$ of the ur-Yukawaon can solely 
appear in the superpotential.

In order to obtain a bilinear relation
$$
Y_e = k \Phi_e \Phi_e ,
\eqno(2.1)
$$
we assume a superpotential term \cite{Koide-O3-PLB08}
$$
W_A= \lambda_A [\Phi_e \Phi_e A_e]+ 
\mu_A [Y_e A_e] ,
\eqno(2.2)
$$
where 
$k=-\lambda_A/\mu_A$ and these fields have U(1)$_X$ charges 
as $Q_X(Y_e)=x_e$, 
$Q_X(\Phi_e)=\frac{1}{2}x_e$ and $Q_X(A_e)=-x_e$.
In addition to the field $A_e$, we introduce a new field $A'_e$  
which couples only to $\hat{\Phi}_e$ as
$[\hat{\Phi}_e \hat{\Phi}_e A'_e]$, and
we also introduce a field $Y'_e$ which composes a mass term 
$\mu^{\prime\prime} [Y'_e A'_e]$ together with $A'_e$ 
similarly to Eq.(2.2).
Since the new field $A'_e$ has the same U(1)$_X$ charge with $A_e$,
we can write the superpotential as follows:
$$
W_A= \lambda_A [\Phi_e \Phi_e A_e]+ 
\mu_A [Y_e A_e] 
+\lambda'_A [\Phi_e \Phi_e A'_e]+ 
\mu'_A [Y_e A'_e] 
$$
$$
+\lambda^{\prime\prime}_A [\hat{\Phi}_e \hat{\Phi}_e A'_e]+ 
\lambda^{\prime\prime\prime}_A \phi_x [Y'_e A'_e] ,
\eqno(2.3)
$$
where $Q_X(A'_e)=-x_e$, $Q_X(\phi_x)=x_\phi$ and
$Q_X(Y'_e)=x_e -x_\phi$.
Here, the reason that we have written 
$\lambda^{\prime\prime\prime}_A  \phi_x$ instead of
$\mu^{\prime\prime}_A$ in Eq.(2.3) is to distinguish $Y'_e$ 
from $Y_e$ in order to prevent $(Y'_e)_{ij}$ from coupling 
with $\ell_i e_j^c$. 
In general, when fields $A_1$ and $A_2$ with the same
U(1)$_X$ charges couple with four terms $\Phi_e \Phi_e$,
$Y_e$, $\hat{\Phi}_e \hat{\Phi}_e $ and $Y'_e$ in Eq.(2.3), 
one of those, for example, 
$[\hat{\Phi}_e \hat{\Phi}_e (c_1 A_1 +c_2 A_2)]$,
can be rewritten as 
$\sqrt{c_1^2+c_2^2}\, [\hat{\Phi}_e \hat{\Phi}_e A'_e]$
without losing generality.
Therefore, the $\lambda^{\prime\prime}_A $-term in Eq.(2.3)
is not an ansatz.
However, the 6th term ($\lambda^{\prime\prime\prime}_A$-term)
in Eq.(2.3) is, in general, given by a linear combination 
of $A_e$ and $A'_e$.
Nevertheless, we have defined $Y'_e$ as the field $Y'_e$ can 
make a mass term only with $A'_e$. 
This is just an ansatz in the present scenario.
From the SUSY vacuum conditions $\partial W/\partial A_e=0$  
and $\partial W/\partial A'_e=0$, 
we obtain the VEV relations (2.1) with $k=-\lambda_A/\mu_A$ 
and 
$$
Y'_e = -\frac{1}{\lambda^{\prime\prime\prime}_A \phi_x } 
\left( \lambda'_A \Phi_e \Phi_e
+\lambda^{\prime\prime}_A \hat{\Phi}_e \hat{\Phi}_e 
+\mu'_A Y_e \right) ,
\eqno(2.4)
$$
respectively.
By substituting Eq.(2.1) for (2.4), we obtain a VEV relation
$$
Y'_e  
= k' (\Phi_e \Phi_e +\xi \hat{\Phi}_e \hat{\Phi}_e) ,
\eqno(2.5)
$$
where
$$
k' =-\frac{1}{\lambda_A^{\prime\prime\prime} \phi_x}\left(
\lambda'_A - \frac{\mu'_A}{\mu_A} \lambda_A \right) , \ \ \
\xi= \frac{ \lambda_A^{\prime\prime} }{
\lambda'_A - \frac{\mu'_A}{\mu_A} \lambda_A } .
\eqno(2.6)
$$
(The other SUSY vacuum conditions $\partial W/\partial Y_e=0$,
$\partial W/\partial Y'_e=0$, $\partial W/\partial \phi_x=0$
and $\partial W/\partial \Phi_e=0$ lead to $A_e=A'_e=0$
for $\phi_x\neq 0$.)

Next, we introduce a field $B_e$ with $Q_X=-\frac{3}{2}x_e +x_\phi$,
and we write a superpotential term
$$
W_B=\lambda_B [\Phi_e Y'_e B_e] .
\eqno(2.7)
$$
The SUSY vacuum condition $\partial W/\partial B_e =0$ 
($W=W_A+W_B$) gives 
$\Phi_e Y'_e=0$, i.e.
$$
\Phi_e (\Phi_e\Phi_e +\xi \hat{\Phi}_e \hat{\Phi}_e) =
(1+\xi)\Phi_e^3 - \frac{2}{3} \xi [\Phi_e]\, 
\Phi_e^2 + \frac{1}{9} \xi [\Phi_e]^2\,  \Phi_e = 0 ,
\eqno(2.8)
$$
from Eq.(2.5).
On the other hand, in general, in a cubic equation
$$
\Phi^3 +c_2 \Phi^2 +c_1 \Phi + c_0 {\bf 1} = 0 ,
\eqno(2.9)
$$
the coefficients $c_i$ have the following relations:
$$
c_2 = -[\Phi] , \ \ c_1 =\frac{1}{2} \left(
[\Phi]^2 -[\Phi \Phi]\right) , 
\ \ c_0= - {\rm det}\Phi .
\eqno(2.10)
$$

The relation for the coefficient $c_2$
$$
c_2 = -\frac{2}{3}\frac{\xi}{1+\xi} [\Phi_e] = -[\Phi_e],
\eqno(2.11)
$$ 
leads to the condition
$$
(\xi+3)[\Phi_e]=0.
\eqno(2.12)
$$
Considering the observed mass spectrum of the charged 
lepton masses, we do not choose a case of 
$[\Phi_e]=v_1+v_2+v_3=0$.
Then the parameter $\xi$ is required as
$$
\xi= -3 .
\eqno(2.13)
$$
[If we take $\xi=-3 +\varepsilon$ ($\varepsilon \neq 0$),
the model will lead to a wrong result $v_1+v_2+v_3=0$.
The value of $\xi$ must exactly be $\xi=-3$.]
The constraint (2.13) puts a strong constraint on the
coefficients $\lambda_A$, $\mu_A$, $\lambda'_A$, $\cdots$,
given in the model (2.3).
Since the constraint (2.13) has been settled by a physical
requirement that the nonzero VEV $[\Phi_e]=v_1+v_2+v_3 \neq 0$
should exist, the parameter $\xi$ is not an adjustable 
parameter in the phenomenological meaning.

From the relation for the coefficient $c_1$, 
we obtain the ratio $R_e$ defined by Eq.(1.5) as follows:
from the coefficient $c_1$,
we have a relation
$$
c_1 = \frac{\xi}{9(1+\xi)}  [\Phi_e]^2 = \frac{1}{2} \left(
[\Phi_e]^2 -[\Phi_e \Phi_e]\right) ,
\eqno(2.14)
$$
so that we can obtain the ratio
$$
R_e \equiv \frac{[\Phi_e\Phi_e]}{[\Phi_e]^2}
= 1-\frac{2\xi}{9(1+\xi)}  = \frac{2}{3} ,
\eqno(2.15)
$$
by using Eq.(2.13).

Although the present model can give a reasonable value of $R_e$,
the cubic equation (2.8) gives $c_0=-{\rm det}\Phi_e =0$, 
which means that the electron is massless, $m_e=0$.
Therefore, next, we are interested in the following ratio
\cite{Koide-PLB662}
$$
r_{123} =\frac{\sqrt{m_e m_\mu m_\tau}}{
(\sqrt{m_e} +\sqrt{m_\mu} +\sqrt{m_\tau})^3}
=\frac{ {\rm det}\Phi_e}{ [\Phi_e]^3} ,
\eqno(2.16)
$$
whose limit $r_{123} \rightarrow 1$ means 
that the electron is massless.
A simple way to obtain a nonvanishing $c_0$
without affecting the values of $c_1$ and $c_2$ 
in the above scenario is to add an ad hoc term
$$
W_{C1}=\varepsilon_1 \lambda_B [\Phi_e][Y'_e][B_e] ,
\eqno(2.17)
$$
to the term (2.7) without violating the U(1)$_X$
symmetry. 
Then, the coefficient $c_0$ is given by
$$
c_0 = \frac{\varepsilon_1 }{1+\xi} [\Phi_e] \left( 
(1+\xi)[\Phi_e\Phi_e]-\frac{\xi}{3}[\Phi_e]^2 \right) .
\eqno(2.18)
$$
By using the relations (2.13) and (2.15), we obtain
$$
c_0=  \frac{1}{6}\varepsilon_1 [\Phi_e]^3 ,
\eqno(2.19)
$$
so that we can obtain the ratio
$$
r_{123} = - \frac{1}{6} \varepsilon_1,
\eqno(2.20)
$$
by recalling the relation $c_0= -{\rm det}\Phi_e$.
If we consider another term
$$
W_{C2}= \varepsilon_2 \lambda_B [\Phi_e Y'_e][B_e] ,
\eqno(2.21)
$$
we can also obtain
$$
c_0 = 3 \varepsilon_2 {\rm det}\Phi_e ,
\eqno(2.22)
$$
where we have used a formula
$$
 {\rm det} A = \frac{1}{3} [A^3]-\frac{1}{2}[A][A^2]
 + \frac{1}{6}[A]^3 .
\eqno(2.23)
$$
Therefore, when we consider both terms (2.17) and (2.21),
$W_C=W_{C1}+W_{C2}$, we obtain
$$
r_{123} = - \frac{\varepsilon_1}{6(1+3\varepsilon_2)} .
\eqno(2.24)
$$
If we assume a traceless field 
$\hat{B_e}\equiv B_e -\frac{1}{3} [B_e]$
instead of $B_e$ in Eq.(2.7), the case corresponds to the
case with $\varepsilon_1=0$ and $\varepsilon_2=-1/3$, and 
we find that $c_0$ 
identically becomes
$c_0 = -{\rm det}\Phi_e$, so that any value of
${\rm det}\Phi_e$ is allowed. 
Therefore, the case is not so interesting.
At present, the parameters $\varepsilon_1$ and 
$\varepsilon_2$ are free, so that we cannot predict
the value of $r_{123}$.

\vspace{3mm}

\noindent{\large\bf 3 \ Concluding remarks}

In conclusion, we have found a superpotential which can lead
to the VEV relation (1.10), $[\Phi_e \Phi_e]=\frac{2}{3} [\Phi_e]^2$.
It should be noticed that, although we have assumed a U(3)
[or O(3)] flavor symmetry in the present paper, it does not
mean that the relation
(1.10) was derived by assuming the symmetry.
The relation (1.10) was obtained by assuming a specific
form (2.3) in the superpotential under the flavor symmetry.
In the superpotential (2.3), the existence of the term
$\Phi_e \hat{\Phi}_e \hat{\Phi}_e$ plays an crucial role
in obtaining the relation (1.10).
If all allowed terms under the symmetry were indiscriminately 
taken into consideration, the model would have become a 
``parameter physics" as well as conventional mass matrix models.
[We have chosen $\xi$ as $\xi=-3$ in Eq.(2.13).
However, as discussed in the previous section (below Eq.(2.13)),
we do not regard $\xi$ as an adjustable 
parameter in the present model.]

Since we have successfully obtained the relation (1.10)
without adjustable parameters,
another problem has risen in the present scenario:
We know that $R=2/3$ is valid only for the charged
lepton masses, and the observed masses for
another sectors do not satisfy $R=2/3$. 
For example, the ratio $R_u$ for the up-quark masses
is $R_u \simeq 8/9$ \cite{Koide-JPG07}.
Can we modify the present scenario as it leads to
$R_u\simeq 8/9$?
At present it seems to be impossible, because there is
no adjustable parameter in the present scenario. 

By the way, on the basis of a Yukawaon model, 
an interesting neutrino mass matrix form 
\cite{Koide-O3-PLB08,Koide-JPG08}
$$
M_\nu \propto \langle Y_e\rangle \left(
\langle Y_e\rangle \langle\Phi_u \rangle
+\langle\Phi_u\rangle \langle Y_e\rangle\right)^{-1} 
\langle Y_e\rangle ,
\eqno(3.1)
$$
has been proposed, where the up-quark mass spectrum 
is given by $\langle Y_u\rangle \propto \langle\Phi_u\rangle 
\langle\Phi_u\rangle$.
The neutrino mass matrix (3.1) can successfully lead to 
a nearly tribimaximal mixing \cite{tribi} under 
an additional phenomenological assumption.
In the successful description of $M_\nu$, it is crucial
that the Majorana mass matrix of the right-handed neutrinos
$M_R$ is given by linear terms of $\sqrt{m_{ui}}$.
Therefore, the bilinear form $\langle Y_u\rangle \propto 
\langle\Phi_u\rangle \langle\Phi_u\rangle$
seems to be valid for the up-quark sector, too.

We also pay attention to the following empirical relation
$$
\sqrt{ \frac{m_{u i}}{m_{u j}} } \simeq 
\frac{m_{e i}+m_0}{m_{e j}+m_0} ,
\eqno(3.2)
$$
where $m_{ui}$ and $m_{ei}$ are masses of up-quarks and
charged leptons.
In fact, for example, the value $m_0=4.36$ MeV gives 
the ratios $(m_e+m_0)/(m_\mu+m_0)=0.0453$ and
$(m_\mu+m_0)/(m_\tau+m_0)=0.0612$ correspondingly to
the observed values $\sqrt{m_u/m_c}=0.0453^{+0.012}_{-0.010}$
and $\sqrt{m_c/m_t}=0.0600^{+0.0045}_{-0.0047}$, respectively.
(Here, we have used quark mass values \cite{q-mass} at $\mu=m_Z$, 
because the quark mass values at a unification scale are highly 
dependent on the value of $\tan\beta=v_u/v_d$.)

These facts suggest a possibility that 
$$
\langle Y_e\rangle \propto \langle \Phi_e\rangle 
\langle \Phi_e\rangle , \ \ 
\langle \Phi_e\rangle \equiv  \langle \Phi_0^e\rangle ,
\eqno(3.3)
$$
in the charged lepton sector, while  
$$
\langle Y_u\rangle \propto \langle \Phi_u\rangle 
\langle \Phi_u\rangle , \ \  
\langle \Phi_u\rangle \propto \langle \Phi_0^u\rangle  
\langle \Phi_0^u \rangle
+\varepsilon {\bf 1} ,
\eqno(3.4)
$$
in the up-quark sector, where ur-Yukawaons $\Phi_0^e$
and $\Phi_0^u$ exactly have the same VEV spectra, but 
the diagonal bases of $\langle\Phi_0^e\rangle$
and $\langle\Phi_0^u\rangle$
are different from each other.

Therefore, there is a possibility that 
all quark and lepton mass spectra 
(in other words, all $\langle Y_f\rangle$)
can be described in terms of only two ur-Yukawaons 
$\Phi_0^e$ and $\Phi_0^u$.
However, in Ref.\cite{Koide-O3-PLB08,Yukawaon}, where 
a supersymmetric Yukawaon model has been investigated
on the basis of an O(3) flavor symmetry, 
the down-quark Yukawaon $Y_d$ has not explicitly been discussed.
In the O(3) model \cite{Koide-O3-PLB08,Yukawaon}, 
since it is assumed that the VEVs of $\Phi_0^e$ and $\Phi_0^u$ 
are real, the observed $CP$ violating phase in the quark sector 
must be inevitably included in the down-quark sector.
Whether such a unified description is possible or not is 
dependent on whether 
a down-quark Yukawaon $Y_d$ can also reasonably be described 
in terms of $\Phi_0^e$ and $\Phi_0^u$.
This will be a touchstone of the Yukawaon approach. 

Finally, we would like to comment on  soft SUSY breaking
terms.
All the results in the present paper have been derived from
using the SUSY vacuum conditions, $\partial W/\partial A_e=0$,
and so on.
We know that SUSY is broken in the realistic world. 
If we add a soft SUSY breaking term, our results will be
changed in principle.
Since we consider that all mass parameters in the superpotential
(2.3) and the VEVs of $Y_e$ and $\Phi_e$ are of the order of 
$\Lambda \sim 10^{15}$ GeV ($\Lambda$ is an energy scale of the
effective theory, and the value $\Lambda \sim 10^{15}$ GeV is
estimated from a Yukawaon model \cite{Koide-O3-PLB08} for the
neutrino sector), the effects will be negligibly small, i.e.
$m_{soft}/\Lambda \sim 10^3 {\rm GeV}/10^{15}{\rm GeV}
\sim 10^{-12}$.  
However, the effect for the conditions (2.13) is troublesome.
In order to keep $\xi=-3$ exactly, we must assume that such
a soft symmetry breaking term which contribute to Eq.(2.13) 
is exactly zero.  
As we have stated in Sec.2, we consider that the constraint
$\xi=-3$ is not a phenomenological one, but a fundamental one
in the model, although, at present, we do not know such a 
reasonable mechanism which keeps $\xi=-3$.
(Or, we may assume that SUSY is unbroken as far as Yukawaons 
are concerned.)
More details for possible soft breaking terms will be 
discussed elsewhere.

\vspace{6mm}

\centerline{\large\bf Acknowledgment} 

This work is supported by the Grant-in-Aid for
Scientific Research, Ministry of Education, Science and 
Culture, Japan (No.18540284).
The author would like to thank Y.~Hyakutake for 
helpful discussions on effects due to soft symmetry 
breaking terms.



\end{document}